# Perspectives to measure neutrino-nuclear neutral current coherent scattering with two-phase emission detector

D.Yu. Akimov[1,2], I.S. Alexandrov[1,2], V.I. Aleshin[3], V.A. Belov[1,2], A.I. Bolozdynya[1], A.A. Burenkov[1,2], A.S. Chepurnov[1,5], M.V. Danilov[1,2], A.V. Derbin[4], V.V. Dmitrenko[1], A.G. Dolgolenko[2], D.A. Egorov[1], Yu.V. Efremenko[1,6], A.V. Etenko[1,3], M.B. Gromov[1,5], M.A. Gulin[1], S.V. Ivakhin[1], V.A. Kantserov[1], V.A. Kaplin[1], A.K. Karelin[1,2], A.V. Khromov[1], M.A. Kirsanov[1], S.G. Klimanov[1], A.S. Kobyakin[1,2], A.M. Konovalov[1,2], A.G. Kovalenko[1,2], V.I. Kopeikin[3], T.D. Krakhmalova[1], A.V. Kuchenkov[1,2], A.V. Kumpan[1], E.A. Litvinovich[3], G.A Lukyanchenko[1,3], I.N. Machulin[3], V.P. Martemyanov[3], N.N. Nurakhov[1], D.G. Rudik[1,2], I.S. Saldikov[1], M.D. Skorokhatov[1,3], V.V. Sosnovtsev[1], V.N. Stekhanov[1,2], M.N. Strikhanov[1], S.V. Sukhotin[3], V.G. Tarasenkov[3], G.V. Tikhomirov[1], O.Ya. Zeldovich[2]

RED (Russian Emission Detector) Collaboration

[1]National Nuclear Research University «MEPhI», Moscow, Russia
[2]SSC RF Institute for Theoretical and Experimental Physics, Moscow, Russia
[3]National Research Centre Kurchatov Institute, Moscow, Russia
[4]Petersburg Nuclear Physics Institute, Gatchina, Russia
[5]Skobeltsyn Institute of Nuclear Physics MSU, Moscow, Russia
[6]University of Tennessee, Knoxville, USA

___

**Abstract**

We propose to detect and to study neutrino neutral current coherent scattering off atomic nuclei with a two-phase emission detector using liquid xenon as a working medium. Expected signals and backgrounds are calculated for two possible experimental sites: Kalinin Nuclear Power Plant in the Russian Federation and Spallation Neutron Source at the Oak Ridge National Laboratory in the USA. Both sites have advantages as well as limitations. However the experiment looks feasible at either location. Preliminary design of the detector and supporting R&D program are discussed.

___

## 1. Introduction

Basic understanding of the Universe more and more requires research under challenging conditions. Often this means searching for extremely rare events that deposit minute amounts of energy. Examples of such research are: search for non-baryonic dark matter, neutrinoless double-beta decay, coherent scattering of reactor neutrinos. Advancing this search requires development of detection technology capable of distinguishing extremely rare and weak signals from large backgrounds resulting from natural radioactivity and cosmic ray interactions.

A two-phase emission detector technology is one of advanced technologies proposed to look for such rare events. Currently it is being developed primarily for detection of dark matter. Recently, collaboration RED (Russian Emission Detector) has been established in order to coordinate joint efforts of Russian institutions in development a large mass liquid Xenon emission detector for neutrino research. In this paper, we examine the possibility to use this technology for a search of still undiscovered neutrino nucleus neutral current coherent scattering off heavy atomic nuclei. We present initial simulations of expected signals in comparison with various backgrounds. We have made some assumptions about detector response at low energies but at the same time we are engaging in the program to measure those parameters at the experiment.

## 2. Neutrino coherent scattering

The process of neutrino elastic interaction with a nucleus via coherent scattering was proposed long time ago [1]. This process has relatively large cross section which can be presented as:

$$\sigma \approx 0.4 \cdot 10^{-44} N^2 (E_\nu)^2 \, \text{cm}^2,$$

where $N$ is the neutron number and the neutrino energy $E_\nu$ is measured in MeV [2]. This formula is valid for neutrino energies up to about 50 MeV, and thus can be applied to reactor, solar and supernova neutrinos. The magnification factor $N^2$ gives a significant increase in cross section for detectors using heavy nuclei as the target. This fact can pave a way for compact neutrino detectors which could have a great impact on nuclear reactor monitoring techniques. This reaction has never been observed experimentally because of the very low energy of recoil nucleus. For example, for neutrinos produced at nuclear power plants the energy of Xenon nuclei recoils are below 1 keV. Recently it has been pointed out that accurate measurement of neutrino coherent scattering can be a sensitive test for the Standard model of electro-weak interactions [3,4]. Requirements for detector are: large mass, high efficiency for sub-keV signals, and capability to achieve extremely low levels of backgrounds. There are several directions where extensive efforts are being directed: low noise Germanium detectors [5], low background NaI detectors [6], and noble liquid emission detectors [7,8,9]. We are exploring the later technology because it provides very low detection threshold, compactness, and good scalability for future detectors of up to 10-20 ton mass.

## 3. Emission detectors

### 3.1. Principle of operation

The emission method of particle detection has been invented 40 years ago [10]. The method allows detection of single ionization electrons, generated in massive non-polar dielectrics such as condensed noble gases and saturated hydrocarbons. In order to amplify signals, electroluminescence of the gas phase excited by drifting electrons can be used. Both original electrons and photons can be used to reconstruct deposited energy, position, time, and to provide particle identification [11]. Such detectors can be constructed to work as "wall less" devices [12] that make them attractive for the use in low background conditions. At present this kind of detector is widely used in experiments searching for cold dark matter in the form of weakly interacting massive particles (WIMPs). Neutrino coherent scattering off heavy nuclei has a similar signature as a WIMP signal. A wall-less detector works as follows (Fig.1):

1) Radiation interacts with the condensed target medium, exciting and ionizing atoms; this process generates a prompt signal that manifests itself in the form of scintillation. This signal serves as a trigger.

2) In response to an applied external electric field, the ionization electrons drift to the surface of the condensed medium where they pass through the surface potential barrier into gas phase. In gas phase, under strong external electrical field they generate an electroluminescent signal (secondary scintillation). An array of photodetectors is used to measure the two-dimensional distribution of the secondary photons and to determine the coordinates of the original event in the plane of the array. Since the secondary scintillation is delayed from the first one by the electron drift time, the third coordinate of the original interaction can be reconstructed from the delay time analyses.

3) From the three-dimensional position reconstruction, a fiducial volume can be defined (*A*, Fig. 1). Then, events originating in the vicinity to the detector walls and associated with radioactive background from surrounding materials can be eliminated. By making the detector sufficiently large and by choosing a target medium with a high stopping power, the fiducial volume can be effectively shielded by the outer detector medium layer (*B*, Fig. 1). Layer *B* can be used as active shield to reject the events in the fiducial volume *A* correlated with interactions detected in the layer *B*. This allows rejection of events associated with multiple scattering of background particles.

4) Analysis of distribution of energy deposited by detected particles between ionization (*EL* signal in Fig.1), and scintillation (*Sc* signal in Fig.1) improves the efficiency of background suppression.

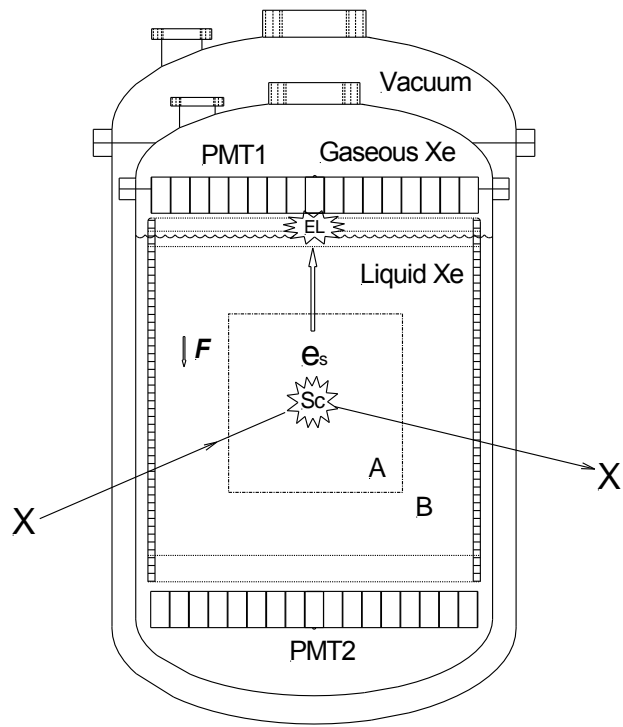

Figure 1. Principal of operation for a "wall-less" liquid xenon emission detector detecting particle *X*: *Sc*− scintillation flash generated at the point of primary interaction between *X* particle and Xe atoms; *EL* − electroluminescence flash of gaseous Xe excited by electrons extracted from liquid Xe by electric field *F* and drifting through the gas at high strength electric field (>1 kV/cm/bar); *PMT1* and *PMT2* − arrays of photodetectors detecting *Sc* and *EL* signals; *A* – the fiducial volume where the occurred events are considered as useful ones; *B* – the layer of liquid xenon used as an active shield for the fiducial volume. The active volume of the detector is surrounded by a highly reflective cylindrical Teflon reflector embodied with drift electrode structure providing uniform electron drift field *F*. The detector is enclosed in a vacuum cryostat made of low-background pure titanium.

The above listed features, along with the availability of super-pure noble gases in large quantities, make condensed noble gases the most attractive media for detection of rare events. For years, technical problems postponed the implementation of emission detector technology into experimental practice. Now, most of these problems are solved [12]: modern gas purification technology permits electron drift length in liquid heavy noble gases of more than 1 meter. The UV light attenuation length can be more than 1 m.

Detection of nuclei with sub-keV kinetic energy is challenging task since there are no available data for the specific ionization yield of nuclear recoils with kinetic energy below 4 keV and for specific ionization yield of electrons/gammas with energy below 15 keV.

In our simulations, we have used the following parameterization for number of ionization electrons produced by nuclear recoils: $12.5 \cdot E^{-0.34}$ $n_{ele}$/keV (where $E$ is measured in keV$_{nr}$) at the energy range above 4 keV (based on compilation of experimental data collected in [13]); and the linear drop to 2 $n_{ele}$/keV at the range from 4 down to 0.1 keV$_{nr}$.

The specific ionization yield of gammas/electrons in the LXe has been assumed to be constant above 20 keV at the level of 23 $n_e$/keV$_{ee}$ and dropping down to 20 $n_{ele}$/keV$_{ee}$ at 15 keV based on the experimental data acquired from [14] and further dropping down to 1 $n_{ele}$/keV$_{ee}$ at 1 keV$_{ee}$.

We are planning to perform a special experimental study of the ionization yield of nuclear recoils at the sub-keV range observing the elastic scattering of a monochromatic beam of neutrons at the IRT MEPhI reactor with a 5-kg LXe two-phase emission detector (RED1) recently used for detection of single electrons [15]. Details of that program are presented in the Section 5 below.

*3.2. Single electron noise*

In two-phase emission detectors, there is observed a specific noise associated with spontaneous emission of single electrons. Experiments performed recently with RED1 brought us to the conclusion [15] that

(1) Single-electron noise is mainly associated with thermal electron emission of ionization electrons accumulated under the liquid-gas interface.

(2) A weak tangential electric field (of the order of several tens of volts per centimeter) promotes a substantial decrease in the intensity of such noise and can be used to improve signal-to-noise ratio in detectors searching for rare events with low energy depositions.

## 4. Detection of neutrino coherent scattering

In this article, we consider two options for the experiment seeking observation of neutrino coherent scattering off atomic nuclei using the nuclear reactor at the Kalinin Nuclear Power Plant (KNPP) in Russia or the Spallation Neutron Source at the Oak Ridge National Laboratory of USA. We plan to build a 100-kg two-phase emission liquid Xe detector, which we call RED100.

*4.1. Neutrino coherent scattering experiment at a nuclear reactor*

A nuclear power plant is an abundant source of electron anti-neutrinos. They are produced in beta decays of radioactive neutron-rich isotopes which are products of fission. In general, antineutrino flux depends on the fuel composition in the core, and therefore, can be used as an independent tool for reactor fuel monitoring. One of the main difficulties in registration of antineutrino from nuclear reactors is associated with neutron background. An additional problem is that industrial reactors are in operation most of the time, and therefore, available time to measure backgrounds is limited. Extra difficulties are caused by the fact that during shutdowns background conditions are different from those of normal operation.

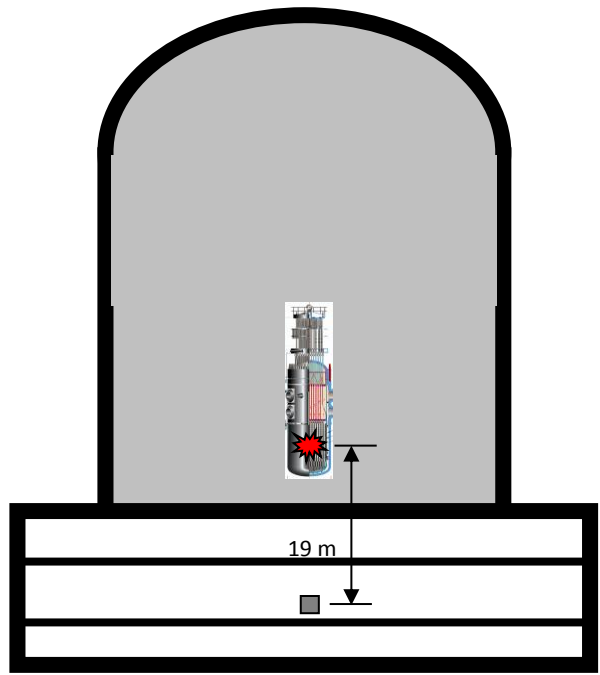

Figure 2. Cross cut of a schematics of the third unit of KNPP. The box shows a potential place for the coherent scattering neutrino experiment with RED100 detector. Distance from the center of reactor core to the center of the detector is 19 m.

As a potential site for the coherent scattering experiment we choose the Kalinin Nuclear Power Plant located near the Udomlia, city of the Russian Federations. This plant is equipped with four WPR-type nuclear reactors with 3000 MW heating power each. The antineutrino detectors GEMMA [16] and DANSS [17] are already installed at one of the units and they have measured the local background conditions at a certain degree of accuracy [18]. A possible room for the coherent scattering experiment is located in the underground gallery below the reactor core as shown in Fig.2.

This location is completely isolated from the operation zone of the reactor. The reactor core together with the reactor shielding and construction elements provides 70 meters of water equivalent (M.W.E.) shielding against cosmic rays in vertical direction and ~ 20 MWE at 60-$70^0$[18]. This reduces the cosmic muon flux by a factor of five and eliminates the hadron component of cosmic rays. The expected rate of cosmic muons passing through the detector is estimated to be about 5 Hz.

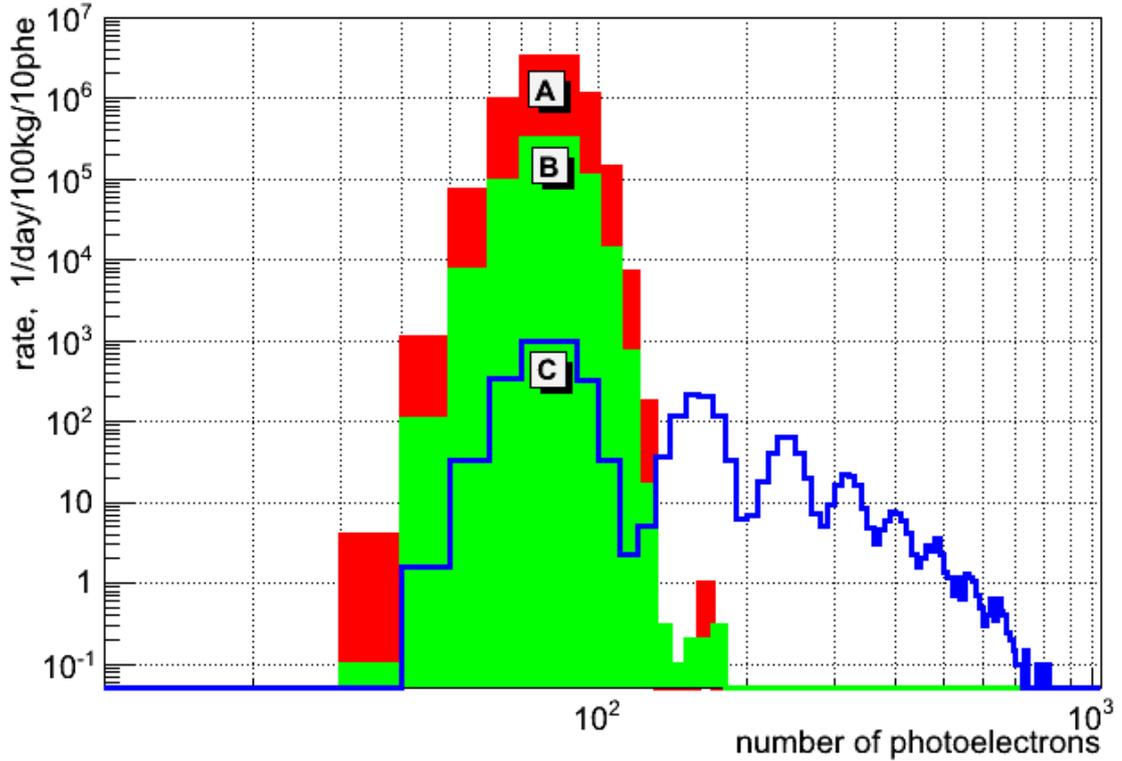

Figure 3. Simulated count rate in the RED100 detector located at 19 meters from the KNPP reactor core associated with the 100 Hz (A) and 10 Hz (B) single emission electron noise and with the neutrino coherent scattering (C).

At this location the distance between the center of the reactor core and the center of the detector is about 19 meters and expected antineutrino flux cross the detector is about $1.35*10^{13} cm^{-2}*s^{-1}$. In the fiducial mass of 100 kg, an expected interaction rate is about ~38000 event per day. However, nuclear recoils produce only a few ionization electrons and single emission electron noise discussed in the Section 3 can be a problem. The results of our simulations for antineutrino signal in comparison with a single electron emission noise at two rates are shown in Fig.3. On the horizontal axis, the number of collected photoelectrons (ph.e) is shown under assumption that each ionization electron generates 80 ph.e. in all PMTs on average due to electroluminescence of the gas phase. On the vertical axis, the rate is expressed in number of events per day per 10 ph.e in 100 kg mass of fiducial volume. Oscillating shape of the curve C is corresponds to detection of one, two and more ionization electrons looking from left to right. For simulations, we select two rates of single emission electron noise of 100 Hz and 10 Hz (shown in red and green, respectively).

It is clearly seen that the spontaneous emission of single electrons accumulated under the interphase surface creates a serious background for antineutrino coherent scattering detection for signals with a magnitude of one ionization electron. In principle this background can be reduced by cleaning the surface of the liquid with a week transverse electrical field. However this requires an additional R&D to prove such a possibility. If we require >2 electrons threshold the rate for antineutrino coherent scattering will be about 433 events per day.

For estimation of backgrounds, we have used a computational model of the detector implemented into GEANT-4 with the following major background sources: the radioactivity of components, the natural radioactivity of external components, neutrons generated by cosmic rays, and $^{85}$Kr uniformly distributed within Xe volume. When we calculated the signals from gammas we assumed electron ionization yield as discussed is section 3.The total mass of Xe in the simulated detector is 200 kg. We use only central fiducial region of 100 kg. This significantly reduces a background because of the self-shielding of Xe.

**Radioactivity of components.** For simulation of background caused by radioactivity of components we have used a compilation of data selected from general databases. An assumed radioactivity for individual components are presented in the Table 1. Radioactivity of Hamamatsu R11410-10 photomultipliers designed for liquid Xe detectors has been measured by the LUX collaboration [19].

Table 1. Assumed radioactivity of the RED100 detector components.

| Component (material) | $^{238}$U | $^{232}$Th | $^{40}$K | $^{60}$Co | $^{137}$Cs |
|---|---|---|---|---|---|
| PMT mBq/unit | 0.4 | 0.3 | 8.3 | 2.0 | |
| Cryostat (Titanium) mBq/kg | 0.2 | 0.25 | 0.93 | | |
| Reflector (Teflon) mBq/kg | 2 | 2 | 15 | 5 | 1 |
| PMT support/heat exchanger (Copper) mBq/kg | 2 | 1 | 4 | 1 | 0.5 |

The estimation of background caused by the decay of $^{85}$Kr with the contamination level accepted from [21] paper shows that, at the region of interest, we can expect no more than 0.5 events per day which is not significant.

**Radioactivity from external environment.** The overall radioactive background near the proposed location at KNPP was measured by the GEMMA collaboration [16]. In the simulations we have used a model of the detector shielded with 100 cm borated polyethylene encapsulated into 10 cm thick iron box. Simulations gave a count rate in the energy range below 1 keV less than 30 events/day from the external radioactivity which is in ten times smaller than the contribution from radioactivity of the internal components.

**Neutrons generated by the nuclear reactor.** Potential place for the detector is well isolated from the reactor hall. Therefore, neutrons originated in the reactor core can't reach the detector. According to the measurement by the GEMMA collaboration at the proposed location there is no difference in the neutron flux between reactor on or off. As a result, we can neglect neutrons generated by the nuclear plant and consider neutrons generated by the cosmic rays only.

**Cosmic rays.** Cosmic rays can produce significant background at a shallow site. However, the cosmic muons deposit quite large energy and therefore do not contribute much in the low energy region. In addition, they and their products will be effectively tagged with a veto system based on plastic scintillation "umbrella" covering the detector and shielding on the top. On the other hand, the neutrons, both prompt and cooled down after number of scatterings in the shielding generated by muons, not passing through the veto system, cannot be vetoed. Those neutrons generate the most dangerous background. For simulations, we assumed that the neutron spectrum has the same energy distribution as that at the Earth surface [20]. We estimate the intensity of neutrons to be attenuated approximately by a factor of five due to the fact that the detector is overburdened under the massive nuclear reactor. This factor of five decreases in muon flux was measured by GEMMA collaboration at the level one step up from the proposed location. Unfortunately, on this stage we can't calculate it precisely because of unknowns in the reactor and shielding structure and we use most pessimistic assumption. However, our estimation is on the save side because of the additional strong reduction of the hardronic components of the cosmic rays. The first approximation on summary of backgrounds caused by the radioactivity of components (A), cosmic neutrons (B) together with the expected signal (C) is shown in Fig. 4. As one can see, all backgrounds are at a level of one order of magnitude or more below the expected signal.

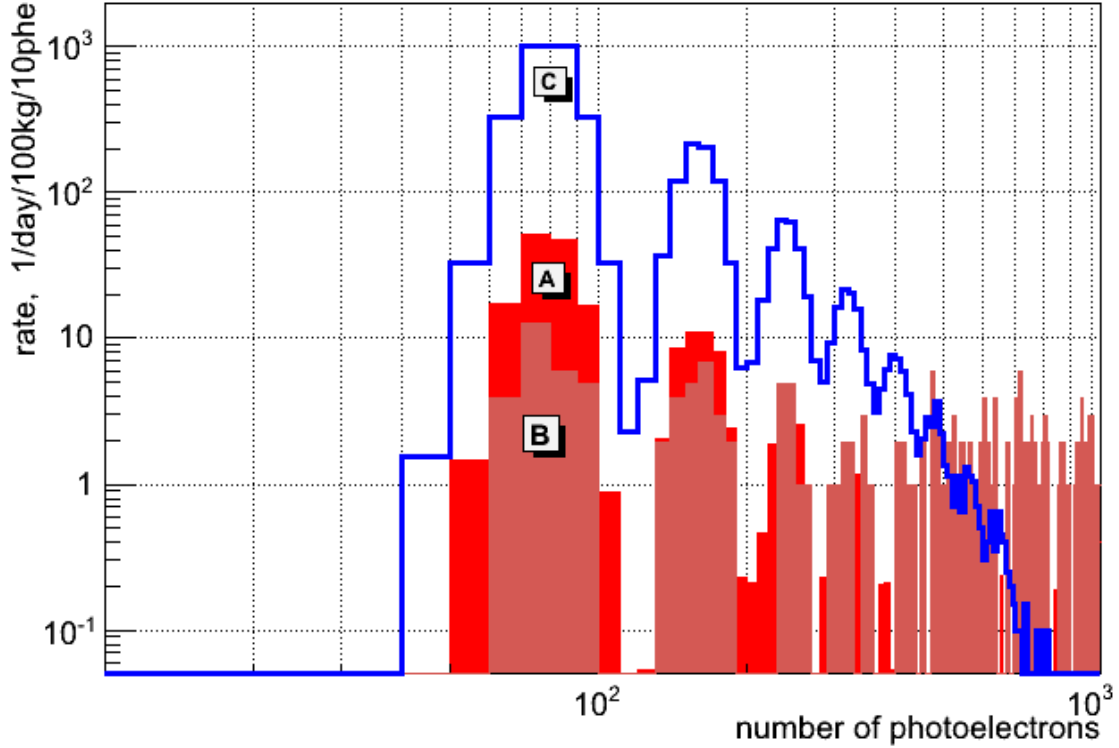

Figure 4. Simulated count rates of backgrounds caused by radioactivity of detector components (A), by neutrons from cosmic rays (B) in comparison with the count rate from the neutrino coherent scattering (C) measured in the RED100 detector located at 19 meters from the KNPP reactor core.

*4.2. Coherent scattering at the stop pion facility*

At stop pion facilities neutrinos produced as a result of decay of pions and muons which are produced by protons interacting on a heavy element target. We consider perspectives to carry out a neutrino coherent scattering experiment at the Spallation Neutron Source (SNS) at the Oak Ridge National Laboratory. The SNS is 1MW and 1 GeV proton accelerator producing neutrons by a spallation on a bulk mercury target. Every proton interacting in the target produces 0.098 $\pi^+$ and 0.061 $\pi^-$. The negative pions are quickly stopped and captured with a little chance to decay. On contrary, the stopped $\pi^+$ and $\mu^+$ decay at rest and create three neutrino species $\nu_\mu$, anti-$\nu_\mu$, $\nu_e$. The proton beam at the SNS hit the target with 600-nsbunches at 60 Hz frequency. Due to the short beam spill the neutrinos at SNS have characteristic time distributions (Fig.5): $\nu_\mu$ from decays of pion produced during a beam spill but anti-$\nu_\mu$ and $\nu_e$ have time distribution with a muon life time of 2.2 usec which provides unambiguous confirmation if positive neutrino signal is detected.

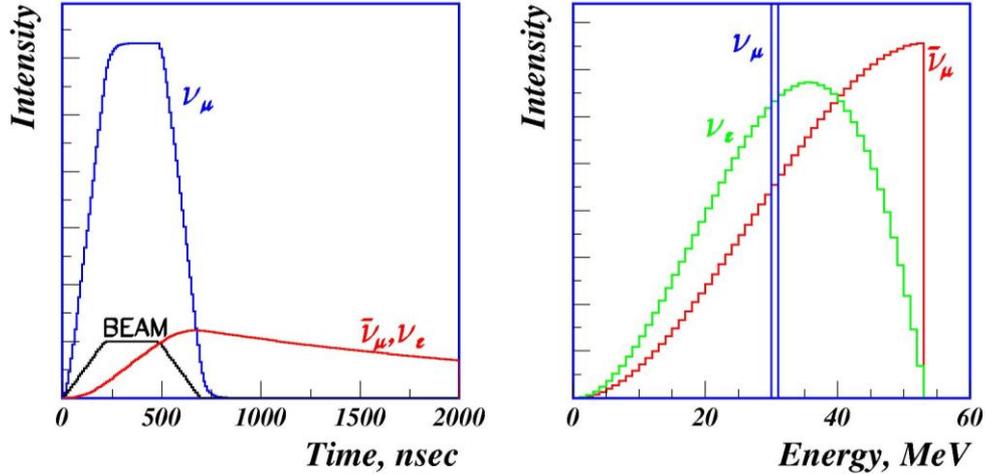

Figure 5. Time and energy distributions for different neutrino flavors produced at the SNS.

The estimated neutrino coherent scattering rate on 100 kg of Xe is 2000 events per year. This is much smaller than the expected rate at the nuclear power plant; however the average recoil energies are much higher. The pulsed beam structure allows measurement of background almost at the same time when the signal is expected to be observed. Expected signal from coherent scattering and spontaneous electron emission is shown on Fig. 6. It is seen that for this particular case loss of efficiency due to the required threshold cut of more than 2 electrons is not significant. Expected rate for signal with three and more electrons is 1400 events/year. For the detector position we assume a location 40 meters from the target, lowered inside 15 meters deep well. Such a well will be an excellent protection from both cosmic induced and SNS generated neutrons.

The big issue is reliable estimation of background caused by a high energy neutrons generated in the target. It is correlated in time with a beam spill and therefore time cut is not really efficient to discriminate from them. Neutron spectra are not easily calculated and calculations are not very reliable (see, for example, [22]).

Due to the SNS duty factor background from cosmic rays and natural radioactivity is suppressed by a factor of 1000. Summary of calculated background for Xe filled detector is shown in Fig.7. Calculations with correction by duty factor gives ~7 events at the region of interest per year from $^{85}$Kr background.

Thus experiment at the SNS has a significant chance for success however more precise knowledge of neutron fluxes at the site is required to make the final conclusion.

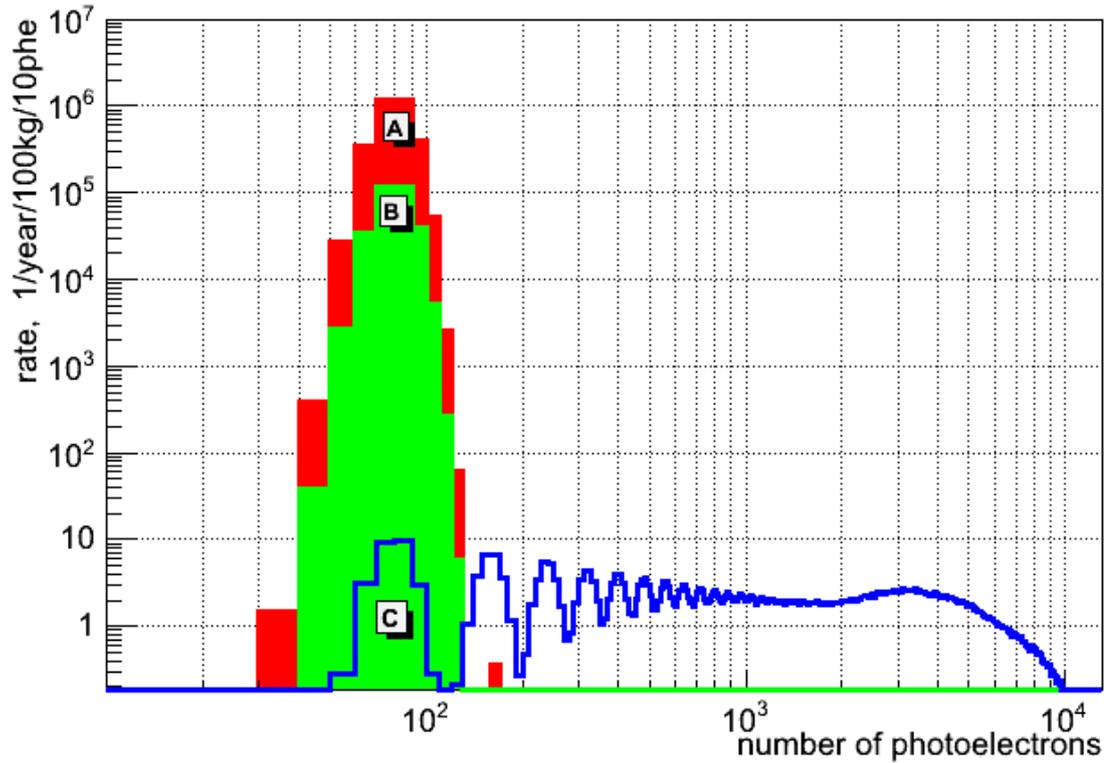

Figure 6. Simulated count rate in the RED100 detector located at 20 m depth in ground and at 40 meters from the SNS target associated with the 100 Hz (A) and 10 Hz (B) single electron emission noise and with the neutrino coherent scattering (C).

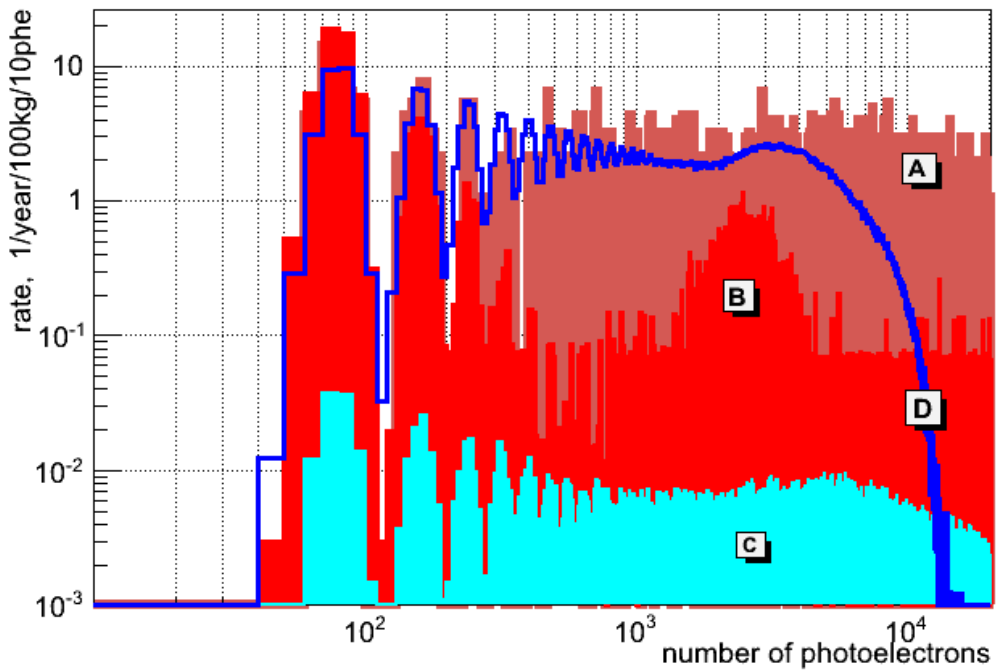

Figure 7. Simulated count rate in the RED100 detector located at 20 m depth in ground and at 40 meters from the SNS target caused by neutrons from cosmic rays (A), by radioactivity of detector components (B), by background neutrons generated by the SNS (C) in comparison with the count rate from the neutrino coherent scattering (D).

## 5. Experiment at IRT MEPhI research reactor

Knowing response of the emission detector in the sub-keV energy range of nuclear recoils is necessary for planning of the experiment to detect the coherent neutrino scattering off atomic nuclei. At the moment, there are no available experimental data below 4 keV for nuclear recoils in liquid Xe.

To measure the response of LXe in the sub-keV nuclear recoil energy range of we are preparing experiment at the 10$^{th}$ horizontal experimental channel GEK10 of the IRT MEPhI research nuclear reactor (Fig.8). The setup includes an interference Fe-Al filter to prepare a quasi-monochromatic neutron beam [23] and RED1 two-phase detector [15].

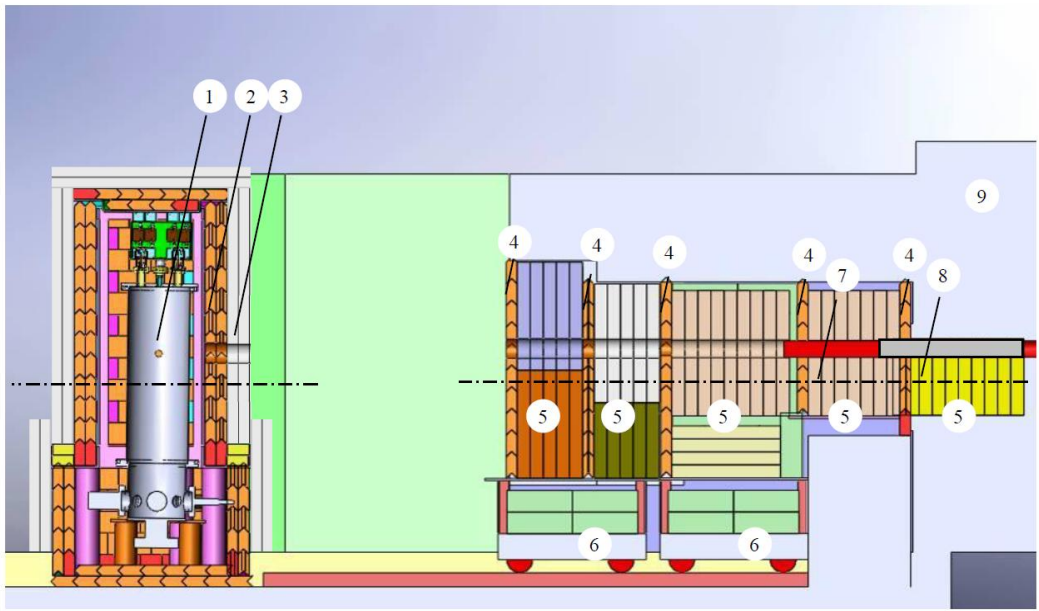

Figure 8. Artistic view of the experimental setup at the research reactor of IRT MEPhI: 1 – two-phase emission detector RED1; 2, 3 – detector passive lead (10 cm) and borated polyethylene (10 cm) shield; 4 – lead layers (5 cm each); 5 – polyethylene blocks made of 5-cm slabs; 6 – paraffin; 7, 8 – iron (30 cm) and aluminum (70 cm) parts of the filter, correspondently; 9 – reactor concrete shielding.

### 5.1. Neutron beam

A quasi-monochromatic neutron beam with the average energy of 24 keV and 3 keV FWHM will be formed by transmitting reactor neutrons through the filter composed of 70 cm long aluminum and 30 cm long iron rods of 10 cm diameter. The iron rod made of a construction steel "steel #3" (in Russian classification) contains as a majority the isotope of iron with mass number 56 that has an interference minimum of the total cross-section at 24 keV. Aluminum is chosen as an additional material because aluminum has resonance peaks in its cross-section at energies above 24-keV, that permits to «cut out» neutrons of higher energy that passed through the interference minima of iron cross section [23]. Spectra of neutrons before and after filter along with cross sections are shown in Fig.9.

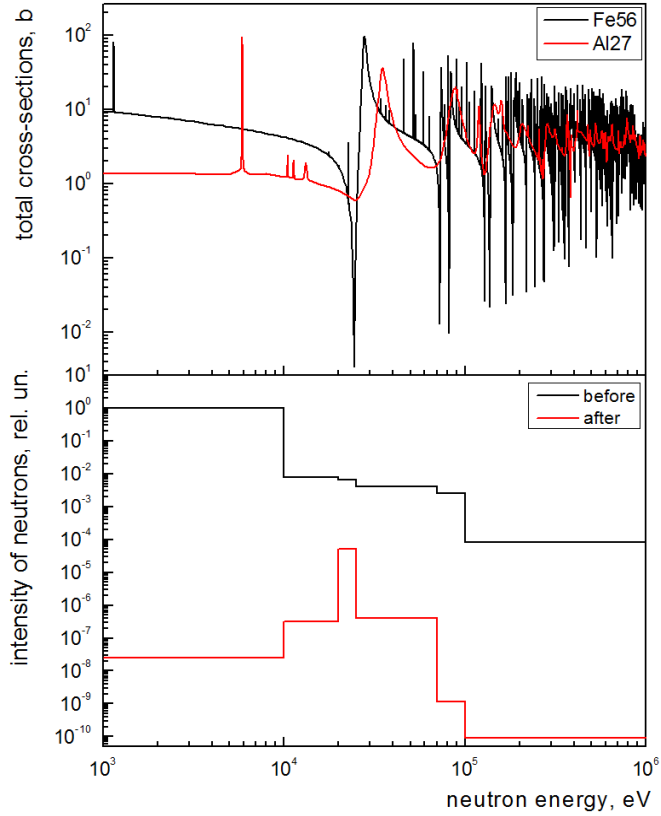

Figure 9. Neutron total cross section of $^{27}$Al and of $^{56}$Fe (upper plots) and simulated neutron beam spectra before and after passing the filter (bottom plots).

Elastic scattering of neutrons with such energy off Xe atomic nuclei will mimic the reactor antineutrino scattering in the detector medium since the energies of the nuclear recoils will be below ~0.7 keV. A filtered neutron beam is formed by a combination of the filter materials and the elements of passive shielding as shown in Fig.8. The passive shielding is made of 32 layers of polyethylene (5 cm) inter-layered with a 1-mm thick Cadmium foil and five 5-cm thick Lead sheets. The Fe/Al filter consists of two 10-cm diameter rods: 30-cm long $^{56}$Fe rod and 100-cm long $^{27}$Al rod installed into a cylindrical hole in the shield as shown in Fig.9. The entire construction (the filter and the shield) is mounted inside a concrete wall of the reactor after the reactor sliding shutter. In the future, we plan to try another filters (such as Silicon containing) that can produce neutrons with higher energies in order to cross check our upcoming results for recoils of a few keV energy in liquid Xenon.

*5.2. Detector RED1*

A detector, which is being used in the experiment at the MEPhI research reactor, has been originally built as a prototype of the ZEPLIN-III dark matter detector [15]. The array of seven PMTs is immersed in the LXe and views the sensitive region (a target) from the bottom through a transparent grid cathode (Fig.10). The electrode structure which contains the target consists of a cathode, an intermediate field-shaping ring, and a mirror anode. There is also a grid electrode in close vicinity to the PMT photocathodes to protect the photocathode chambers from the strong electric field produced by the cathode. This electrode is kept at the same potential as the photocathodes. The thickness of the LXe layer is 22 mm, the diameter of the target is 105 mm, the 5-mm wide gas gap between the liquid surface and the anode where electroluminescence is generated. All gaps between the electrodes are located in the liquid Xe; this minimizes the probability of discharge between electrodes. The potential difference between the anode and the cathode is up to 15 kV (already tested). The HV is supplied to the electrode system via Ceramaseal made high voltage feedthroughs. The amount of LXe in the target is 0.6 kg while the total amount required to fill the chamber is ~5 kg.

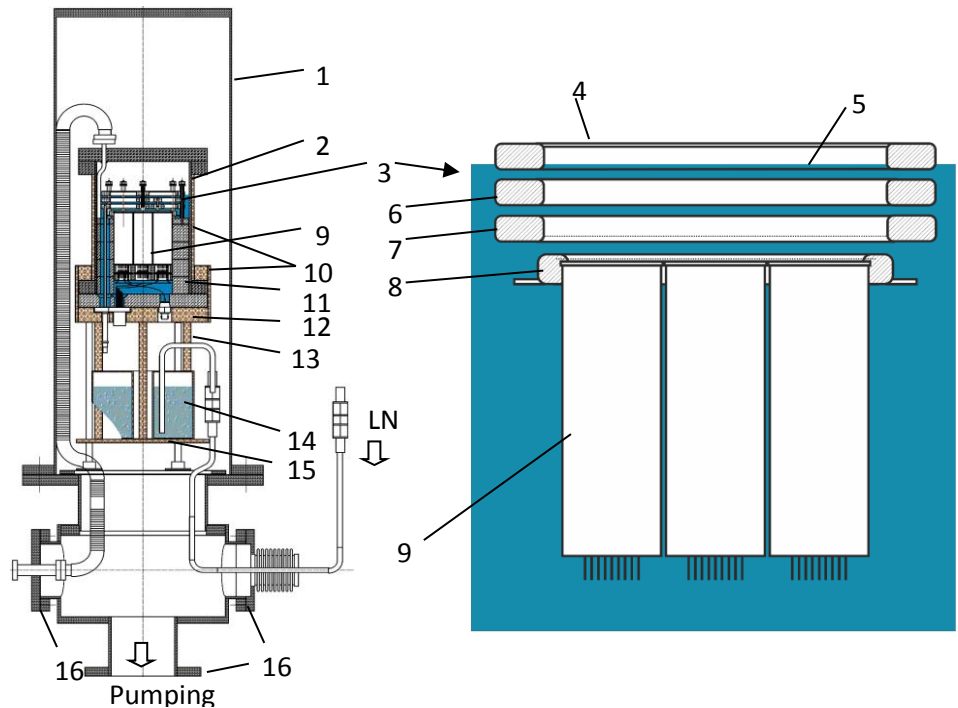

Figure10. RED1 emission detector for measurement of the LXe response at the sub-keV energies of nuclear recoils. 1 – vacuum jacket; 2 – chamber for LXe filling; 3 – electrode structure; 4 – mirror anode; 5 – free surface of LXe; 6 – intermediate field shaping ring; 7 – wire cathode (0.1 mm diam., 1 mm pitch); 8 – screening electrode (0.1 mm diam., 1 mm pitch); 9 – PMTs; 10 – copper jacket; 11 – LXe displacers; 12 – copper base; 13 – copper fingers; 14 – $LN_2$ vessel; 15 – copper base of $LN_2$ vessel; 16 – vacuum ports.

The VUV-sensitive $MgF_2$-windowed photomultipliers of the FEU-181 type made by MELZ (Moscow) are used to detect the 175-nm xenon scintillation light. This type of the PMT has a 30-mm multialkali photocathode. The measured quantum efficiency of the photocathode at 200 nm is 15%, and there is no dependence within 5% accuracy of the signal when temperature varied from the room value to -120°C. The PMT array has common divider placed outside the chamber.

With this configuration, we have a photoelectron yield for one ionization electron extracted from the liquid to the gas phase of 15 at $U_c$= -10 kV and $U_a$ = 1.85 kV. Later we plan to upgrade the existing detection system by installation of additional array of 7 PMTs in the gas. The anode will be made from mesh at this case. The Monte Carlo simulation shows that about 30 phe per one electron can be expected with such geometry.

Note that the detector RED1 can be used to study the LAr response: the $MgF_2$-windowed PMTs are sensitive to the Ar emission at 125 nm, and the multialkali photocathode can operate at a temperature of LAr environment (T = -185 °C).

*5.3. Ionization response measurement*

Measurement of the specific ionization yield will be done by comparison of the obtained charge spectrum with the expected one where this value will be used as a free parameter. The energy distribution of the nuclear recoils from elastic scattering of neutrons with 24-MeV energy is nearly flat with a maximum energy of ~ 0.7 keV for Xe.

*5.4. Scintillation response measurement*

A scintillation response of the LXe for the nuclear recoils has been measured down to 2 keV by several experimental groups (see compilation of the experimental data in [13]). For lower energies, it could be estimated by integral method. The entire concept of this method is to sum waveforms for the large number of events having the charge signal of $N_e \geq 2$ (the trigger must be tuned to the EL signal in this case). In this energy range, scintillation signals for nuclear recoils having sub-keV energies are expected at the level of single photoelectrons only. The region of the cumulative waveform before the sum EL signal will be populated by single photoelectron events; part of them will be the noise (PMT single photoelectron noise plus the weak light from scintillations in LXe outside the sensitive region of the detector), and another part is single photoelectrons from the events of neutron scattering, which produced a trigger. The signals of second type enhance the single photoelectron population in the time interval corresponding to the full drift time of electrons before the EL signal. To obtain the scintillation response one must subtract the first component which can be estimated from the part of waveform before this interval.

This method has been already tested for the energies $E_{nr}$>4 keV, and the result was found to be in agreement with those obtained with other methods [13].

## 6. Conclusion

We are starting a long term program with the goal to detect of neutrino coherent scattering off atomic nuclei with a two-phase xenon detector. Beyond the great impact on fundamental physics and verification of the Standard Model with observation of the neutrino coherent scattering [24] the project leads to development of a relatively compact and effective detector of nuclear reactor antineutrinos. Such a device could be used for

- monitoring the output power of an industrial reactor [25, 28];
- monitoring reactor fuel burnup [26, 27];
- diagnostics of critical situations [29].

**Acknowledgements**

This work was supported by the RF Government under contracts of NRNU MEPhI with the Ministry of Education and Science of №11.G34.31.0049 from October 19, 2011 and № П881 from May 26, 2010 and, in part, by the Russian Foundation for Fundamental Research under the contract of №11-02-00668-a and by the Russian Ministry of Education and Science via grants 8174, 8411, 1366.2012.2.

The authors gratefully acknowledge the logistical and technical support and the access to infrastructure provided by the research reactor IRT MEPhI staff, in particularly, Chief engineer Dr. Alexander Portnov and NRNU MEPhI administration. Our special thanks to Dr. Alexander Starostin of ITEP for many enlightening discussions about background conditions at the KNPP and Prof. William Bugg of the University of Tennessee for careful reading of manuscript and many useful comments.